\documentclass[preprint2]{aastex63}

\submitjournal{ApJ}

\shorttitle{The birth of radio-loud quasar 013815+00}
\shortauthors{Kunert-Bajraszewska et al.}

\begin{document}


\title{Caltech-NRAO Stripe 82 Survey (CNSS). IV. The Birth of Radio-loud Quasar 013815+00}

\correspondingauthor{Magdalena Kunert-Bajraszewska}
\email{magda@astro.uni.torun.pl}

\author{Magdalena Kunert-Bajraszewska}
\affiliation{Institute of Astronomy, Faculty of Physics, Astronomy and Informatics, NCU, Grudziadzka 5/7, 87-100, Torun, Poland}

\author{Aleksandra Wo\l{}owska}
\affiliation{Institute of Astronomy, Faculty of Physics, Astronomy and Informatics, NCU, Grudziadzka 5/7, 87-100, Torun, Poland}
    
\author{Kunal Mooley}
\affiliation{National Radio Astronomy Observatory, Socorro, New Mexico 87801, USA}
\affiliation{Cahill Center for Astronomy, MC 249-17, California Institute of Technology, Pasadena, CA 91125, USA}

\author{Preeti Kharb}
\affiliation{National Centre for Radio Astrophysics(NCRA) $-$ Tata Institute of Fundamental Research(TIFR), S.P. Pune University Campus, Post Bag 3, Ganeshkhind, Pune 411007, India}

\author{Gregg Hallinan}
\affiliation{Cahill Center for Astronomy, MC 249-17, California Institute of Technology, Pasadena, CA 91125, USA}

\begin{abstract}
It is believed that the gas accretion onto the supermassive black holes (SMBHs) is the main process of powering its luminous emission, which occurs in optical, UV and X-ray regimes and less frequently in radio waves. The observational fact that only a few percent of quasars are radio-loud is still an unresolved issue concerning the understanding of the active galactic nucleus (AGN) population. Here we present a detection of a rapid transition from the radio-quiet to the radio-loud mode in quasar 013815+00 (z=0.94) which coincides with changes of its UV-optical continuum and the low ionization MgII broadline.
We interpret this as an enhancement of accretion onto a central black hole of mass about $10^9$ solar masses. As a consequence a new radio-loud AGN was born. Its spectral and morphological properties indicate that it went through the short gigahertz-peaked spectrum (GPS) phase at the beginning of its activity and has now stabilized its flux density at the level of a few mJy. The radio morphology of 013815+00 is very compact and we predict that with such short-term jet activity its development will be very slow. The observed luminosity changes of the accretion disk are shorter than the lifetime of the new radio phase in 013815+00.

\end{abstract}



\section{Introduction} \label{sec:intro}
The radio emission of active galactic nuclei (AGNs) is dominated mainly by synchrotron radiation of electrons accelerated by powerful jets. However, such AGNs traditionally called radio-loud constitute only a small part of the entire population. About 90\% of AGNs are much fainter in the radio and they are
called radio-quiet sources \citep{kellerman1989}.
The origin of their radio emission is still unclear and widely debated as well as the nature of the event that sparks the radio-loud phase. In general, source classification for radio-loud/radio-quiet is a matter of the definition used and since it is based on selected observational properties it can be somewhat ambiguous \citep{kharb2014, foschini2017}. The presence or absence of a strong relativistic jet seems to be a better criterion for differentiating these two groups of objects \citep{padovani2017}. However, it has its limitations since it cannot be used in the case of unresolved sources with faint radio flux densities. 

According to the generally accepted model \citep{fanti1995, readhead1996, O'deabaum} the radio sources having a peak in their radio spectra above 1 GHz represents the early stages of the life cycle of radio-loud sources. These are the compact ($<$1 kpc) Gigahertz-peaked spectrum (GPS) and High-frequency peaked (HFP) sources. The latter group are sources with peaks above 5 GHz \citep{dallacasa2000}, but the peak value may vary, and that is why in the literature these objects are often referred to as GPSs. The peak emission frequency is inversely proportional to the source size \citep{O'deabaum} showing that the next in this development sequence are compact steep spectrum (CSS) sources. CSSs have larger linear sizes ($\sim 1-20$kpc) and peak frequencies below 1 GHz. The ability to perform sensitive observations at low radio frequencies has recently revealed many objects with spectral peaks observed at frequencies far below 1 GHz called Megahertz-peaked spectrum (MPS) sources \citep{callingham2017, keim2019}.
The final stage of the life cycle is a large scale ($>20$ kpc) radio-loud AGN of Fanaroff-Riley I (FRI) or II (FRII) morphology \citep{fr1974}.

Recent reports show that the GPS and CSS sources account for only 2\% of the brightest AGNs \citep{sotnikova2019}. But in the case of low luminosity objects the compact population is much larger and even makes up the majority of samples tested \citep{sadler2014, baldi2015}. This excess of small AGNs in relation to "adult" objects gave rise to the suggestion that not all compact young sources evolve into extended objects. According to one of the hypotheses the radio activity is a short-term phenomenon for many AGNs lasting $10^4 - 10^5$ years and an AGN may undergo numerous such short phases during its lifetime \citep{rb97,czerny2009, kunert2010, an2012, wolowska2017}. The reason for this premature termination of activity could be, e.g. instabilities of the accretion disk that cause the jet disruption \citep{czerny2009}.  
However, there are also results indicating that at least for some low luminosity AGNs their compact radio morphology  does  not  mean  a  young  age but rather suggests a presence of different fuelling mechanisms \citep{hardcastle2019, capetti2019}.
Regardless of the hypotheses presented above, the GPS phase seems to be the starting point for the development of each new phase of radio activity and is associated with the emergence of a new radio jet. In this sense a young GPS source means ongoing episode of the accretion disk outburst. 

\begin{figure*}[t]
\centering
\includegraphics[scale=0.42]{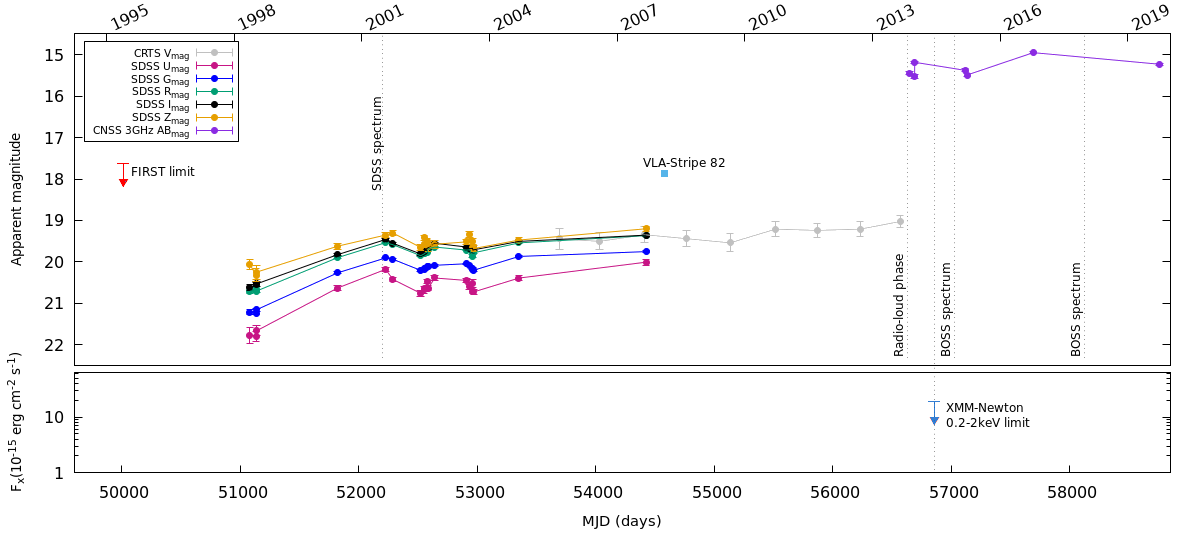}
\caption{Top panel: Long term optical light curve for SDSS J0138+0029. The SDSS photometric measurements in the {\it u',g',r',i',z'} filters are shown with small filled circles. The three epochs of SDSS spectroscopy as well as the date of radio activity burst are indicated by the dotted lines. The yearly median and rms of the unfiltered CRTS (Catalina Transient Survey) measurements are shown with gray filled circles. The radio measurements have been converted to apparent magnitude on the AB system of \citet{oke1983} and they are: the FIRST 3$\sigma$ rms noise level (red arrow), the VLA-Stripe 82 source emission (blue square), the 3 GHz light curve detected by CNSS and later 2016 and 2019 VLA measurements (purple circles). Bottom panel: shows the upper limit of the X-ray flux from XMM-Newton observations, which amounts to  $F_{0.2-2 keV} < 18.0 \times 10^{-15}$~erg~cm$^{-2}$~s$^{-1}$ \citep{lamassa2013}.}
\label{photometry}
\end{figure*}

The first detection of switched-on radio activity was in a z=1.65 quasar, VTC233002-002736, found in the pilot Caltech-NRAO Stripe 82 Survey (CNSS) carried out over a 50 $\rm deg^2$ region of Stripe 82 \citep[CNSS Paper I;][]{Mooley}. We reported a GPS-like characteristic of its radio spectrum, typical for a young source. 
The burst of radio activity has been also discovered recently in radio-quiet narrow-line Seyfert 1 (NLS1) galaxies \citep{lahteenmaki2018}. This is interpreted as the launch of a relativistic jet indicating the similarity in the radio behaviour of NLS1 galaxies and more energetic AGNs. 

In this paper we present the first few years of development of the radio-loud phase of another quasar, CNSS J013815+002914 (hereafter 013815+00). 
The radio emission of 013815+00 was first discovered in the CNSS survey, on 2013 December 20 at 3 GHz.
We report multi-epoch and multi-frequency observations of this source in \S\ref{sec:obs_data}. 
We found that the rapid changes of radio spectrum of quasar 013815+00 
transforming the initial GPS shape into a flat spectrum are associated with the changes of brightness in the AGN accretion disk. The high-resolution VLBA observations confirm that the burst of radio activity is associated with the appearance of a small radio jet. 
We discuss these findings and implications in detail in \S\ref{sec:discussion} and summarize our work in \S\ref{sec:summary}.

Throughout this paper we assume a cosmology in which $\rm H_0= 70~km~s^{−1}~Mpc^{−1}$,$\rm \Omega_m= 0.3$ and $\Omega_{\lambda}= 0.7$. The radio spectral index $\rm \alpha$ is defined in the sense $\rm S \propto \nu^{\alpha}$.

\vspace{2cm}
\section{Observations and data analysis}
\label{sec:obs_data}

\subsection{Radio data }
The radio emission of quasar 013815+00 (RA 01h38m15.061s, Dec 00d29m14.07s) was detected on 2013 December 20 at 3 GHz by the Caltech-NRAO Stripe 82 Survey (CNSS), a multi-epoch radio transient survey carried out between December 2012 and May 2015 with the Jansky VLA \citep{Mooley}. It was designed for systematically exploring the radio sky for slow transient phenomena on timescales between one day and several years. 

All the calibration process of these observations was done using a custom-developed, semi-automated AIPSLite/Python-based pipeline developed at Caltech and
CASA\footnote{http://casa.nrao.edu} software. 
The observed 3 GHz flux density of quasar 013815+00 was 2.41$\pm$0.09, 2.25$\pm$0.09, 3.07$\pm$0.09, 2.56$\pm$0.09 and 2.31$\pm$0.09 mJy in the five epochs of observations indicating relatively stable behaviour (Fig.\ref{photometry}).

\begin{table*}[ht]
\caption{Line and continuum measurements from SDSS/BOSS spectra}
\centering
\begin{tabular}{lccc}
\tableline\tableline
\multicolumn{1}{c}{QSO 013815+00}&
\multicolumn{1}{c}{2001}&
\multicolumn{1}{c}{2015}&
\multicolumn{1}{c}{2018}\\
\tableline
MgII flux [$\rm 10^{-17} erg~s^{-1} cm^{-2} $] & 145$\pm$15 & 356$\pm$6 & 307$\pm$8\\
MgII FWHM [$\rm km~s^{-1}$]& 9428$\pm$1283 & 9964$\pm$257 & 9857$\pm$386\\
MgII $\rm log_{10}[L_{MgII}/erg~s^{-1}$]& 42.81$\pm$0.05 & 43.20$\pm$0.01 & 43.14$\pm$0.01\\
Cont. flux density at 3000$\rm \AA$ [$\rm 10^{-17} erg~s^{-1} cm^{-2} \AA^{-1}$]& 5.0$\pm$0.9 & 7.1$\pm$0.4 & 5.2$\pm$0.5 \\
$\rm M_{BH}$ [$\rm 10^9 M_{\odot}$]& 1.6$\pm$0.7  &  2.3$\pm$0.2 & 1.8$\pm$0.3\\
$\rm L_{bol} [10^{44} erg~s^{-1}]$& 35.8$\pm$6.4& 50.6$\pm$2.9&36.7$\pm$3.6\\
$\rm [O III]$ $\lambda$5007 flux [$\rm 10^{-17} erg~s^{-1} cm^{-2} $] &  $ - $ &  42$\pm$2 & 30$\pm$2\\
$\rm [O III]$ $\lambda$5007 FWHM [$\rm km~s^{-1}$]& $ - $ & 658$\pm$30 & 658$\pm$69  \\
$\rm [O III]$ $\lambda$5007 $\rm log_{10}[L_{\lambda 5007}/erg~s^{-1}$]& $ - $ & 
42.27$\pm$0.02 & 42.13$\pm$0.03 \\
\tableline
\end{tabular}
\label{table}
\end{table*}

After the discovery of radio quasar 013815+00 we performed its follow-up wide range observations with the JVLA in A configuration on 2016 November 16 (project VLA/16B-047, 6 hours), using five receivers covering the spectrum from 1000 to 16884 MHz (L, S, C, X, Ku). The observing setup was the correlator with 16 spectral windows and 64 2-MHz-wide channels. Another epoch of observations was obtained in 2019 November (project  VLA/19B-209, 6 hours) in D configuration using four receivers (S, C, X, Ku) and the same observing setup. Then the detailed calibration and imaging of VLA data was carried out using CASA.

The high-resolution observations of 013815+00 were made with the VLBA at C-band on 2016 February 26. The data were correlated at the Array Operations Center in Socorro (USA).
We phase-referenced our observations to the VLBA calibrator 0137+012. We used the strong source J0237+2848 as fringe finder and bandpass calibrator. In order to carry out the spectral analysis of the radio components of our source we divided the available bandwidth at C-band receiver into two subbands centered at 4.5 and 7.5 GHz. This strategy allowed us to obtain images at two frequencies in one scan. Data reduction (including editing, amplitude calibration, instrumental phase corrections and fringe-fitting)
was performed with the standard procedure using the NRAO AIPS\footnote{http://www.aips.nrao.edu} software.
After this stage the AIPS task IMAGR was used to produce the final total intensity images. The flux densities of the main components of the target source were then measured by fitting Gaussian models using AIPS task JMFIT. The linear size of the source was calculated based on the largest angular size measured in the 4.5 GHz image contour plot.

In order to complement the spectrum of 013815+00 at low frequencies, sub-GHz observations with upgraded GMRT in Band 3 and Band 4 were carried out on 2018 March 16 and 19. Unfortunately, Band 3 data were corrupted, probably due to incorrect phase calibration. However, good quality Band 4 (560-810 MHz) data were obtained, and analyzed with CASA software. In order to obtain as many measurement points as possible, while maintaining sufficient signal-to-noise (S/N) ratio, the data were divided into two spectral windows, each one processed separately and added to the radio spectrum plot (Fig. \ref{spectrum}).

Finally, in order to characterize the significant changes in the radio spectrum of 013815+00 it has been fitted with modified power-law model \citep{snellen1998}:\\

\begin{equation}
\footnotesize
\mathrm{S(\nu)=\frac{S_{p}}{(1-e^{-1})}\times(\frac{\nu}{\nu_{\mathrm{p}}})^{\alpha_{\mathrm{thick}}}\times(1-e^{-(\frac{\nu}{\nu_{\mathrm{p}}})^{\alpha_{\mathrm{thin}}-\alpha_{\mathrm{thick}}}}) }\\
\end{equation}

where $\rm \alpha_{thick}$ is the optically thick spectral index, $\rm \alpha_{thin}$ is the optically thin spectral index, and $\rm S_{p}$ and $\rm \nu_{p}$ are the peak flux density and peak frequency, respectively.
Since the GMRT observations were carried out in a different time than the VLA observations, only VLA data were fitted in each epoch. However, the low frequency GMRT data are also indicated on the plot. The fits characterize the significant changes that occur in the optically thin part of spectrum of quasar 013815+00. At lower frequencies, as might be expected, the spectrum is more stable. It is also very likely that there are no changes at GMRT frequencies. Therefore, the location of these points in relation to the modeled VLA radio spectrum curve influenced the choice of the best model. 
In addition, to obtain a satisfactory fit to the VLA data we had to limit the number of free parameters by one of the spectral indices.

\subsection{Optical observations}
\label{sec_opt}
To track the brightness changes in the optical and UV range of quasar 013815+00 we have collected all available photometric data points since 1998 untill the end of 2013 from the Sloan Digital Sky Survey (SDSS) and Catalina Sky Survey (CRTS; \citet{drake2009}. We averaged the CRTS data in segments of 1 year and put the yearly median and RMS values on Fig. \ref{photometry} together with the SDSS data. 

Additionaly, three epochs of SDSS spectroscopic measurements of 013815+00 are available and were analyzed in this article. The first spectroscopic observation was made by the SDSS on 2001 October 21 over the wavelength range of 3800-9200 $\AA$ at a spectral resolution of $\simeq$2000. The other two were made by the Extended Baryon Oscillation Spectroscopic Survey (eBOSS) over the slightly different than previous wavelength range 3600-10000 $\AA$ at a spectral resolution of $\simeq$2000 on 2015 December 8 and 2018 January 6.

The SDSS spectra were corrected for Galactic extinction with the reddening map of \citet{schlafly2011}, and shifted to the rest-frame wavelength by using the SDSS redshift.
The decomposition of the spectra has been done using the IRAF package and assuming the following components: power law (representing the emission from an accretion disk), FeII pseudo-continuum, and suite of Lorentzian and Gaussian components to model the broad and narrow emission lines. In order to remove the contribution from FeII emission, the iron template from \citet{Bruhweiler2008} was fitted to both spectra. The template was convolved with Gaussian profile with different dispersion values for kinematic broadening of FeII lines, in order to find the most accurate one. The emission lines were fitted with Lorentzian/Gaussian function (with single or multiple components, depending on a line) to determine fluxes and FWHMs. The FWHM of [O III] line quoted in Table \ref{table} was obtained by fitting a single Gaussian to enable comparison of this value with those obtained for other compact radio objects. The quality of the spectra did not allow us for more complex modeling of the [O III]$\lambda\lambda$4959,5007 doublet.

The error estimation of the continuum flux density was made using the rms method. 
The uncertainty of the line flux measurements has been estimated using the standard formula for noise averaging
$\rm \sigma_f = \sigma_c L / \sqrt{N}$, where $\rm \sigma_c$ is the rms of continuum flux density at 3000$\AA$, $L$ is the integration interval, and $N$ is a number of spectrum samples.
The error of the line width has been calculated by finding minimum and maximum widths of the Lorentz/Gaussian line fit at which the integral of the fit changes by $\rm \pm\sigma_f$ keeping the amplitude of the fit fixed.
The properties of the broad and narrow lines resulting from the fit to each spectrum are listed in Table \ref{table}.

The black hole mass was estimated from MgII$\lambda$2800 line, using the following relation \citep{trakhtenbrot2012}:\\

\begin{equation}
\footnotesize
\frac{\mathrm{M_{BH}}}{\mathrm{M_{\odot}}}=5.6\times10^6 \left( \frac{\lambda L_{3000}}{10^{44} ~\mathrm{erg~s}^{-1}} \right)^{0.62}
\left[ \frac{\mathrm{FWHM(MgII)}}{10^3~\mathrm{km~s}^{-1}} \right]^2  
\end{equation}

\section{Discussion}
\label{sec:discussion}
The radio emission of quasar 013815+00, which is located in the Stripe82 region, was detected on 2013 December 20 at the level of few mJy with VLA at 3 GHz. 
Previous surveys of this part of Stripe82 were carried out in 1995 and 2008 with VLA at 1.4 GHz under the names: the Faint Images of the Radio Sky at Twenty-Centimeters (FIRST) \citep{white97} and the VLA survey of the SDSS Southern Equatorial Stripe (VLA-Stripe 82) \citep{hodge2011}, respectively. Both surveys did not report a detection of this quasar at the catalog detection limit at the source position of 0.78 mJy (FIRST) and 0.67 mJy (VLA-Stripe 82). 
However, a careful inspection of the radio images, revealed an unresolved weak emission at the location of 013815+00 amounting to 0.26$\pm$0.06 mJy in the case of more sensitive VLA-Stripe 82 observations. 
The radio image rms noise level at the position of our quasar amounts to 0.106 mJy (FIRST) and 0.084 mJy (VLA-Stripe 82). This indicates that in 2008 the emission of the 013815+00 quasar was at the 3$\rm \sigma$ rms noise level (0.25 mJy) of VLA-Stripe 82 survey. After the burst of radio activity, the latest measurement of its 1.4 GHz flux density amounts to 1.61$\pm$0.11 mJy and means about a six-fold increase in flux density at radio waves and a significant change in the value of radio-loudness parameter. To calculate the last we adopted radio-loudness definition from \citet{kimball}: $\rm logR = (M_{radio}-M_{i})/-2.5$,
where $\rm M_{radio}$ is a radio absolute magnitude at 1.4 GHz and $\rm M_{i}$ is a Galactic reddening corrected i-band absolute magnitude. For the detection level of 0.26 mJy (from VLA-Stripe 82) the value of radio-loudness parameter amounts to $\rm logR = 0.6$ and increases after the radio activity ignition to $\rm logR = 1.4$ crossing the boundary $\rm logR = 1$ above which the source is considered to be radio-loud.

\begin{figure}[t]
\centering
\includegraphics[scale=0.39]{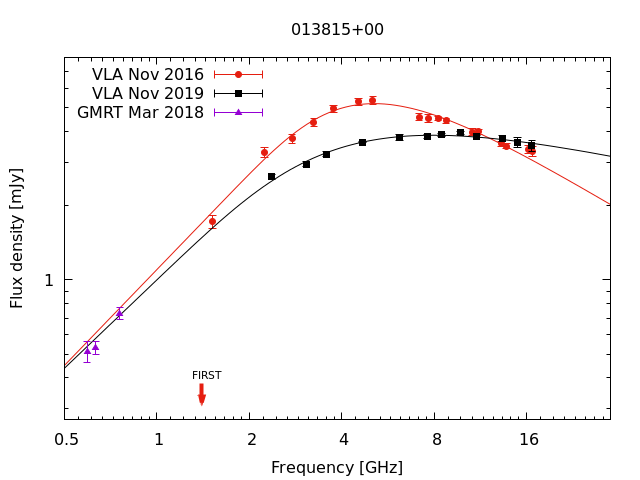}
\caption{Radio spectrum of quasar 013815+00 consisting of Very Large Array measurements in the range 1-20 GHz and low frequency GMRT observations in the range 0.6-0.8 GHz. The red arrow indicate the 3$\rm \sigma$ upper limit at 1.4 GHz from the FIRST survey.}
\label{spectrum}
\end{figure}

\subsection{Analysis of radio properties}
The follow-up observations of quasar 013815+00 performed with the VLA (1-15 GHz) on 2016 November 16 revealed a radio spectrum peaking at $\rm \nu_{p}=4.72\pm0.19\, GHz$ with a flux density $\rm S_{p}=5.17\pm0.07\, mJy$ and the spectral index $\rm \alpha_{thin}= -0.75$ and $\rm \alpha_{thick}=1.3\pm0.1$ of the optically thin and thick part of the spectrum, respectively (Fig. \ref{spectrum}). 
The 5 GHz luminosity of this object at this epoch of observations has a moderate value of $\rm log_{10}[L_{5 GHz}/W~Hz^{-1}]=25.4$. 

\begin{figure*}[ht]
\centering
\includegraphics[scale=0.42]{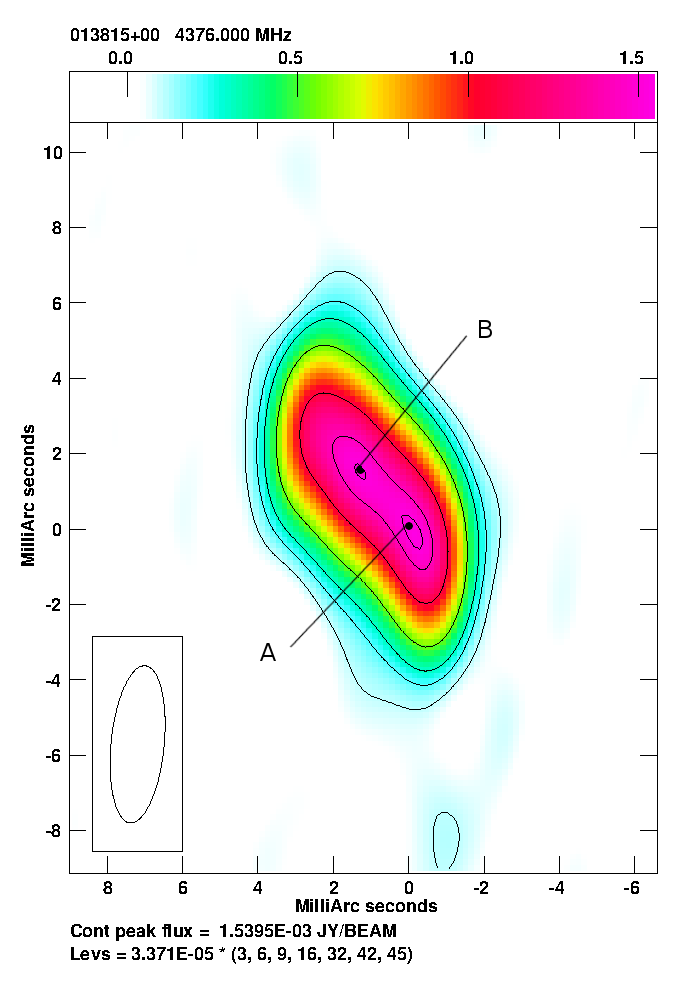}
\includegraphics[scale=0.42]{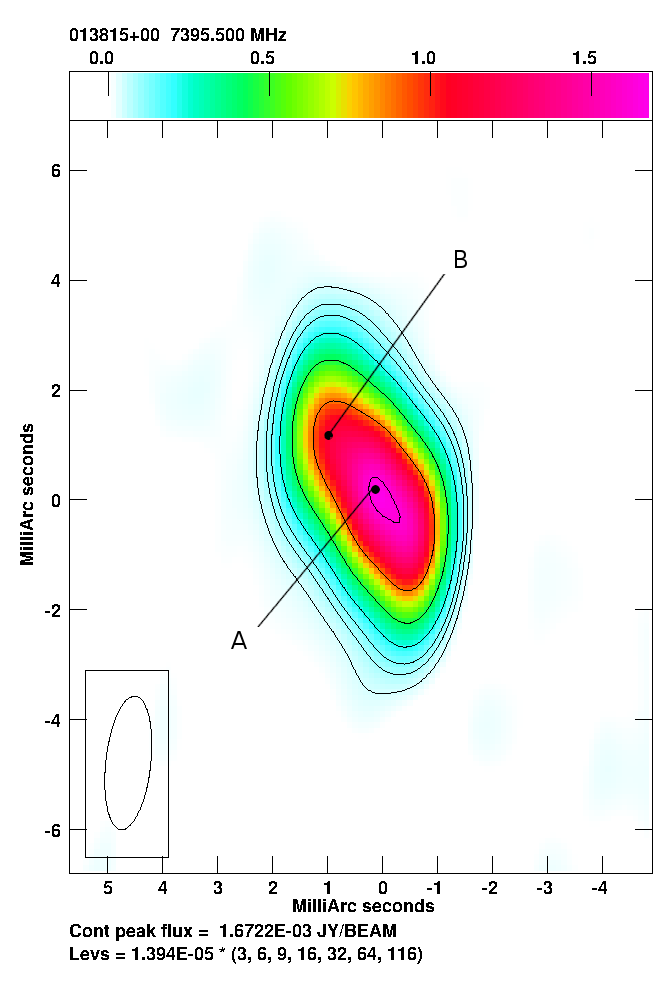}
\caption{Very Long Baseline Array 4.5 and 7.5 GHz images of quasar 013815+00. The first contour level corresponds to $\rm \approx 3\sigma $ and the radio beam size is indicated in the lower-left corner. The color bar shows the scale of intensity of radio emission in mJy. The two identified radio components are indicated with dots.  }
\label{radio_images}
\end{figure*}

The VLBA high-resolution observations at 4.5 and 7.5 GHz of this source performed on 2016 February 26 show a slightly resolved radio structure of the total projected linear size $l \rm \approx67 ~pc$. The radio morphology indicates the presence of two compact features: (A) the central one with flux densities 2.14$\pm$0.07 and 2.02$\pm$0.03 mJy, and (B) the northeastern component with flux densities 3.02$\pm$0.09 and 2.80$\pm$0.03 mJy at 4.5 and 7.5 GHz, respectively (Fig. \ref{radio_images}). The radio spectral index of both components between 4.5 and 7.5 GHz is flat and amounts to $-0.11$ and $-0.15$ for the central and northeastern component, respectively. We suggest that the radio structure of 013815+00 is a core-jet type with the central component being the radio core.

Over the next three years the radio morphology of quasar 013815+00 must undergo a rapid change, which is reflected in the significant evolution of its spectrum. The new VLA spectrum obtained on 2019 November 10 shows a drop in luminosity in wide frequency range and spectral flattening of the thin part of the spectrum. The spectrum peaks at $\rm \nu_{p}=4.21\pm0.20\, GHz$ with a flux density $\rm S_{p}=3.51\pm0.07\, mJy$ and the spectral index $\rm \alpha_{thin}= -0.27\pm0.03$ and   $\rm \alpha_{thick}=1.19$ of the optically thin and thick parts of the spectrum, respectively (Fig. \ref{spectrum}). We interpret the 2016 GPS spectrum of quasar 013815+00 as a burst of new radio jet activity. As the radio jet expands, its flux density decreases and the spectrum starts to be dominated by the compact radio core.
Further changes in the spectrum are still possible and only long-term monitoring will allow us to finally determine the shape of the spectrum. Current observations show that the source after a radio-quiet phase will stabilize its radio flux density at the level of a few mJy.

\begin{figure}[t]
\centering
\includegraphics[scale=0.38]{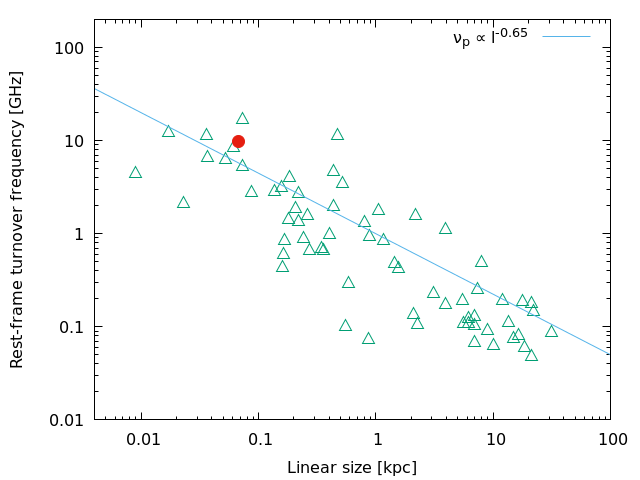}
\caption{Rest-frame turnover frequency versus projected linear size relation for CSS and GPS sources studied by \cite{O'deabaum}. The GPS quasar 013815+00 is indicated with a red dot.}
\label{plot_odea}
\end{figure}

The radio properties of quasar 013815+00 in its first phase of activity, namely the convex spectrum and compact morphology, make it follow the anticorrelation found by \citet{O'deabaum} which relates the spectral turnover frequency (the frequency at which the spectrum peaks) in GPS and CSS sources to their projected linear sizes (Figure \ref{plot_odea}). There is a continuous distribution of these objects across the $\nu_{p} - l$ plane, which suggest that GPS and CSS sources are scaled versions of each other. However, the long-term monitoring of samples of GPS sources lasting 10-20 years shows that with time even the majority of them cease to adhere to the generic GPS source properties \citep{dallacasa2000,torniainen2005, orienti2006, sotnikova2019}.
This may be due to the fact that the GPS phase occurs in many, or even all, populations of radio sources and is what at the initial stage of development connects them with each other. In fact, however, these groups may not be related. In particular, this may apply to the division into galaxies and quasars. According to \citet{snellen1999} GPS quasars can be a subclass of flat-spectrum quasars. Another complication in this model of the life cycle of radio source are the results from low luminosity AGNs, which show that these sources have different lifetime distribution \citep{hardcastle2019, capetti2019}.

013815+00 shares several properties similar to GPS quasars like convex spectrum and compact core-jet morphology, but has lower luminosity and its spectrum evolves rapidly. We thus conclude that 013815+00 is a newborn radio source with a very short transient GPS phase. We suggest that the duration of the GPS phase may scale with the source luminosity and thus with the power of the jet. 

Using the relation between the jet power and radio luminosity at 1.4 GHz discussed by \citet{rusinek2017} we estimated the power of the newborn jet in 013815+00 to be $\rm log_{10}[P_{j}/erg~s^{-1}]=44.1$. 
This value is intermediate, and it falls within a wide range of jet power values for young AGNs \citep{wojtowicz2020}.
In turn, such a wide distribution of jet power values may mean the existence of different subclasses of sources within the group of young radio sources \citep{berton2017, fan2019}, as we have already mentioned above.

\subsection{Photometric and spectroscopic changes}

The collected photometric observations of quasar 013815+00 from the period of 15 years show an over of 1 magnitude brightening in all SDSS bands during the first three years of observations (Fig. \ref{photometry}). At the peak of this brightening the first spectroscopic measurement of the source has been made by the SDSS on 2001 October 21. Over the next twelve years the quasar light curve shows a typical AGN submagnitude variability and slow but systematic increase in brightness of another $\sim$0.5 magnitude in V-band of CRTS observations. 

\begin{figure*}[ht]
\centering
\includegraphics[scale=0.41]{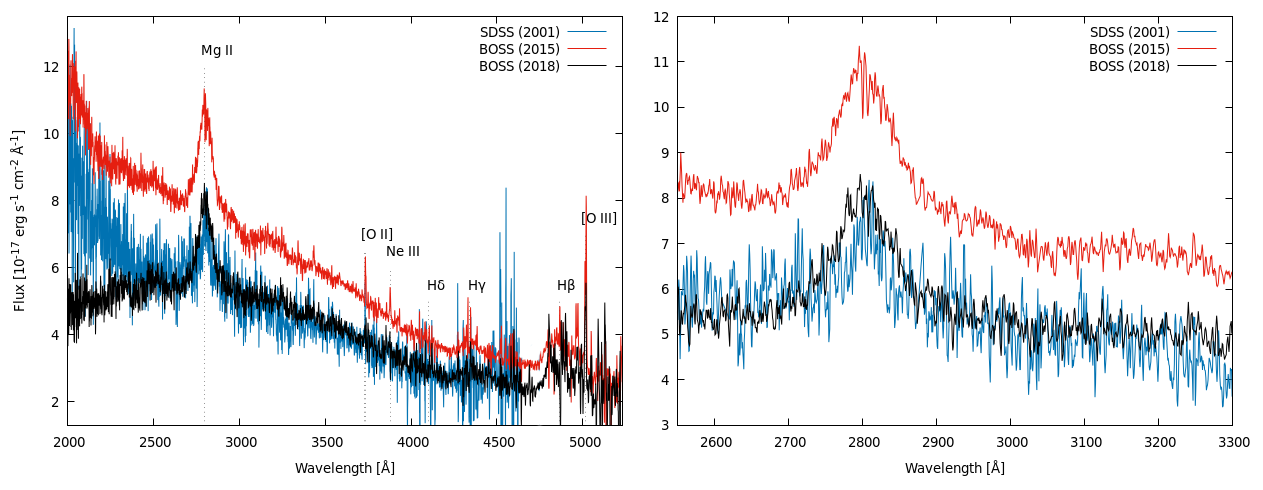}
\caption{Left panel: shows the SDSS 2001 (blue) and BOSS 2015 (red) and 2018 (black) spectra corrected for telluric absorption and with line identifications.
The right panel: shows the MgII line profile changes. 
The spectra were smoothed with a 3-pixel boxcar.}
\label{spectroscopy}
\end{figure*}

The other two spectroscopic observations were made by the eBOSS on 2015 December 8 and 2018 January 6. They took place after the burst of 013815+00 radio activity in 2013 and enabled us to follow the changes in the continuum and the brightness of the emission lines that happened after this event. 
An increase of the level of continuum emission is clearly visible when comparing 2001 and 2015 spectra and amounts to 1.4 (Figure \ref{spectroscopy}).
We are not able to give a precise date of the beginning of this growth. However, the increased emission of the source in the optical-UV range continues for the next two years after the new radio jet launching in 2013.
In period 2015-2018 the luminosity drops to the level close to that of the 2001 observations. During this time we also observe a significant change in the radio spectrum of quasar 013815+00 which we interpret as a dissipation of jet energy as a result of its expansion.

\subsubsection{MgII$\lambda$2800 line behaviour}
The value of redshift of 013815+00 means that the Balmer emission lines are not in the SDSS (2001) spectrum but the MgII$\lambda$2800 line is clearly present. The BOSS (2015, 2018) spectra also contain magnesium line and several other emission lines, such as [O III]$\lambda$5007, but the H$\alpha$ line is out of range.
All this makes our comparative analysis of the quasar spectral properties limited mostly to the magnesium line. The value of the full-width at half-maximum (FWHM) of MgII line measured in all spectra (Table 1) indicates that it is a strong broad emission line typical for unobscured AGN. 
However, the flux of MgII line shows great variability over these 19 years which indicates that it has responded to the continuum changes, although the scale of theses responses differ between epochs. Both the continuum and the MgII line flux increased significantly between 2001 and 2015 (Table 1). Over the next two years, the continuum decreased and in 2018 it reached the level observed in 2001 while the MgII line did not change that much. The flux of the MgII line showed only a slight decrease but still remained at a higher level compared to 2001. 
This probably implies a time lag with which the MgII line reacts to continuum changes. We could not notice this for the continuum growth phase due to the lack of spectroscopic observations between 2001 and 2015. The time delay between the continuum variations and the response of the Mg II emission line has been already reported for radio-quiet \citep{czerny2019} and radio-loud objects \citep{nalewajko2019}.
What is more the very recent results on the study of a large sample of extremely variable radio-quiet AGNs \citep{homan2019} indicate that indeed the MgII line does respond to the continuum changes, although the scale of behaviour variability is very broad. In most of the studied AGNs the MgII line varies less than the continuum.
In the case of the quasar 013815+00 the increase of the MgII line flux observed in 2015 is greater than the change in the continuum emission, which indicates a more prominent response to the continuum changes than in typical AGN.
According to \citet{guo2019} the MgII line flux changes can be well explained as a
photoionization response to extreme continuum variability. However, a wide range of possible responses among AGNs may indicate additional factors that affect the change in the MgII line \citep{popovic2019, homan2019,yang2019}.

In the case of radio-loud AGNs, the complex behavior of the MgII line described above further complicates the presence of a radio jet. 
There are detailed studies showing correlation between the jet kinematics and continuum and emission line variability in quasars and blazars suggesting that the jet affects the gas producing the broad emission lines and continuum \citep{leon2013, berton2018, chavushyan2020}. The scale of the nonthermal contribution may depend on the jet viewing angle and on the jet power.

In order to estimate the contribution of the jet emission to the total optical$-$UV emission in the case of the quasar 013815+00 we used the method proposed by \citet{shaw2012}.  
We calculated the value of nonthermal dominance $\rm NTD=L_{obs}/L{p}$ parameter, where $\rm L_{obs}$ is the observed continuum luminosity and $\rm L_{p}$ is the predicted continuum luminosity estimated based on the continuum-line luminosity correlation found by \citet{shen2011} for a nonblazar sample.
NTD parameter values calculated for both spectra (2015 and 2018) of the radio-loud phase of our quasar are very similar and do not exceed 1 ($\rm NTD < 1$). This may suggest that the observed continuum and MgII emission in the radio-loud phase is reflecting the contribution from the accretion disk only.

\subsubsection{[O III]$\lambda$5007 line properties}
Similarly, the radio jets of compact radio sources are thought to be strongly interacting with the denser phases of the interstellar medium (ISM) in the near-nuclear regions of their host galaxies \citep{holt2008}. Such strong jet-cloud interactions can cause the enhancement of the [O III]$\lambda$5007 line emission as well as the radio emission boosting \citep{labiano2008, morganti2011}. Additionally, the emission line kinematics of the CSS/GPS objects are more extreme in terms of line shifts and widths than the general populations of nearby AGN and radio galaxies with extended radio sources \citep{holt2008, liao2020}.
In the case of quasar 013815+00 only the spectroscopic BOSS (2015, 2018) observations allow us to 
characterize its [O III]$\lambda$5007 line emission and follow its changes after the burst of radio activity. 
Unfortunately the low ($\rm S/N<5$) for [O III]$\lambda\lambda$4959,5007 doublet prevented us from carrying out a more complex modeling of this emission.

The values of [O III]$\lambda$5007 line luminosity (Table 1) and 5 GHz radio luminosity from both epochs make the 013815+00 follow the correlation between emission line luminosity and radio power of radio-loud AGNs \citep{kunertlabiano2010, morganti2011}. This probably does not indicate any extra radio emission rising from the jet-cloud interactions but places the source among the so-called high-excitation galaxies (HEGs) on the plot. The value of $\rm L_{[O III]}$ is also the highest one when comparing it to the [O III] emission line luminosities of radio-quiet `changing-look AGNs' reported by \cite{macleod2016}
which suggests that radio-loud 013815+00 produces more ionizing photons.
The FWHM of [O III]$\lambda$5007 of 013815+00 is also larger than the FWHM distribution peak for extended radio sources ($\rm \sim 200-300~ km~s^{-1},$) and is in the range of values reported by \citet{holt2008} for CSS/GPS sources.
This may imply some impact of the jet on the narrow emission line region (NLR) kinematics ($<$1~kpc) in quasar 013815+00. However, according to the high-resolution radio observations the radio jet of quasar 013815+00 is small and the whole radio structure extends to tens of parsecs, which significantly limits the area of its impact unless the gas producing [O III] emission is located closer to the central black hole. Such an explanation has been discussed in the literature and the recent Hubble Space Telescope study of nearby Type 2 quasars shows that quasars with a more compact [O III] morphology have broader nuclear emission lines probably as a result of a gas concentration near the AGN \citep{fischer2018}. This gas may be driven out at the later states by radio jets or AGN-winds, which may also operate on larger scales (100$-$1000s of pc).  Without more detailed observations of emission region in quasar 013815+00 we are unable to distinguish any of the interpretations cited.

\subsection{Characteristic of the accretion process}

We used the measurements of MgII emission line and luminosity at 3000$\rm \AA$  to estimate the black hole mass of quasar 013815+00 (sec. \ref{sec_opt} and Table 1).
The obtained values for all epochs are consistent within the measurement errors. For further estimations we use the average black hole mass from the three epochs.
In the next step we calculated the bolometric AGN luminosity (Table 1) in its radio-quiet stage using $\rm \lambda L_{3000}$ and a conversion factor of 5.3 to convert from monochromatic to bolometric luminosity \citep{runnoe2012}. With the Eddington luminosity defined as $\rm L_{Edd} = 1.3 \times 10^{38}~M_{BH}$ and averaged black hole mass, we find the Eddington ratio ($\rm \lambda_{Edd} = L_{bol}/L_{Edd}$) of 0.015. If we estimate $\rm L_{bol}$ using the value of $\rm \lambda L_{3000}$ from the BOSS (2015) spectrum after the source went into the radio-loud phase and again calculate the $\rm \lambda_{Edd}$, its value increases to 0.021. However, during the next two years the luminosity and hence the Eddington ratio drops and again amounts 0.015, which is the same as in the radio-quiet phase.

The general formula defining bolometric luminosity says that $\rm L_{bol}$ is directly proportional to mass accretion rate $\rm \dot{M}$ (in units of $\rm M_{\odot}/yr$) and the radiative efficiency $\rm \eta$ of the accretion process: $\rm L_{bol}= \eta \dot{M}c^{2} $ \citep{coziol2017}.
Since the value of the BH mass of 013815+00 probably does not change after the radio activity ignition we conclude that the observed increase in the Eddington ratio ($\rm \sim \eta\dot{M}c^{2} / M_{BH} $) is a result of higher accretion rate and/or radiative efficiency. 
The change in the bolometric luminosity is not big (by a factor of 1.4) compared to the typical factor of 4 obtained for changing-look radio-quiet AGNs by \citet{macleod2019}. 
Still, the relatively small change of the accretion rate and/or radiative efficiency that took place in quasar 013815+00, led to the launching of a new radio jet and a significant change in radio luminosity from $\rm log_{10}[L_{1.4GHz}/W~Hz^{-1}]=24.1$ to 24.9.
Interestingly, the recent studies of the shape of the radio luminosity function \citep{malefahlo2019} show change of its behaviour in the luminosity range $\rm log_{10}[L_{1.4GHz}/W~Hz^{-1}]=24 - 25$ what is interpreted as a change of the dominating population from radio-quiet to radio-loud. 

The estimated values of the black hole mass and Eddington ratio of 013815+00 are in agreement with those calculated for other young AGNs, although the range of these values for AGNs considered young is very wide \citep{berton2017, liao2020}. The accretion power of 013815+00 is larger than its jet power, which implies low jet efficiency $\rm P_j/L_{bol}\sim 0.02$ and makes the source comparable to flat-spectrum radio quasars (FSRQ) and high-excitation FRII objects \citep{rusinek2017, fan2019, wojtowicz2020}.
The large black hole mass of 013815+00 may suggest that the AGN is old and it probably has already gone through many such phases of enhanced accretion. We have only captured one of these phases. 
Perhaps this behaviour could also be compared to the high/hard-to-soft state transition observed in black hole binaries and very recently also for a sample of young AGNs \citep{wojtowicz2020}, with the relativistic jet being launched in a high state.

\section{Summary}
\label{sec:summary}
The burst of radio activity in quasar 013815+00 detected on 2013 December 20 changed the source status from radio-quiet to radio-loud. The expanding new radio jet is responsible for the convex radio spectrum peaking at $\sim$5 GHz in the early stage of activity but also for the rapid changes observed in this spectrum during the next few years. Ignition of radio emission coincides with a significant increase in the brightness of the accretion disk and emission lines. However, within the next two years, the disk brightness returns to its original state, which shows how fast changes occur in this new source. We suggest that the accretion disk behavior and birth of the radio source are the result of an enhancement of the accretion rate and/or radiative efficiency. The large black hole mass of 013815+00 may indicate that such periods of increased accretion may have already occurred in the past for this object.

At gigahertz frequencies, the new radio source can currently be classified as young flat-spectrum quasar. 
It went through a short GPS phase comparing to more powerful objects and its radio flux density has now stabilized at the level of a few mJy. We predict that with such short-term activity of relatively weak radio jet, the development of the radio structure will be very slow in this quasar and it may remain compact for most of its life.

\acknowledgments
We thank the anonymous referee for helpful  suggestions that led to improvement of the paper. We are grateful to Agnieszka Ku\'zmicz and Bo\.zena Czerny for the useful discussion.
The National Radio Astronomy Observatory is a facility of the National Science Foundation operated under cooperative agreement by Associated Universities, Inc. We thank the staff of the VLBA and VLA for carrying out these observations in their usual efficient manner.
This work made use of the Swinburne University of Technology software correlator, developed as part of the Australian Major National Research Facilities Programme and operated under licence.
MKB and AW acknowledge support from the  „National Science Centre, Poland” under grant no. 2017/26/E/ST9/00216.
KPM is a Jansky Fellow of the National Radio Astronomy Observatory.
\software{CASA (\citet{McMullin2007}, AIPS (\citet{vanMoorsel1996}, IRAF (\citet{Tody1986,Tody1993})}



\end{document}